# Time domain analysis of bunched e-beam tapering-enhanced superradiance at zero-slippage conditions

Amir Weinberg[1*], Leon Feigin[2], Ariel Nause[2], Avraham Gover[1†]

[1]Center for Light-Matter Interaction, Tel Aviv University, Ramat Aviv, Israel

[2]Department of Physics, Ariel University

* amirwe@tauex.tau.ac.il

†gover@eng.tau.ac.il

Exceptionally, a bunched electron beam can get trapped by its own spontaneously emitted synchrotron undulator radiation in the self-interaction scheme of TES (Tapering Enhanced Superradiance). In this scheme, excess radiative energy is extracted from the beam by tapering the undulator after the bunch trapping. To avoid slippage of the radiation wavepacket away from the bunch, the interaction takes place in a waveguide with a slow group velocity mode. Here we study the time domain waveform of the Electric Field and the radiation energy buildup in this scheme. We compare its analytical theory and numerical simulations in an ideal exemplary setup based on a rectangular waveguide and a planar undulator

## 1. Introduction:

Bunched beam superradiance (SR) is the coherent spontaneous emission of free electrons that radiate mutually in phase into a radiation mode [1]. Consequently, the field amplitudes of the wavepackets of the individual electrons interfere to enhance the resultant field in proportion to the number of interacting electrons N, and the radiative energy of the emitted radiation grows in proportion to $N^2$. This feature is the hallmark of Dicke's superradiance [2] that was discussed originally in the context of coherent emission from atomic dipoles.

There are two kinds of bunched-beam SR:

1. Emission by a continuous or a long pulse of electrons which is bunched at an optical frequency $\omega_b$ within the spontaneous emission frequency line of the beam radiation. In this case the coherent SR emission is narrow, centered around $\omega_b$.

2. Emission by a single short bunch electron beam. In this case the electrons emit coherently with each other in a wide frequency band as long as the duration of the bunch $\sigma_t$ is shorter than an optical period of the emitted radiation:

$$\omega\sigma_t \ll 2\pi \qquad (1)$$

In this case the coherent SR emission is enhanced, but its spectrum is determined by the spontaneous emission bandwidth of the single free electron-light interaction process, which in our case is Synchrotron Undulator Radiation (or the spontaneous emission process of FEL).

SR emission can be enhanced by external injection of a radiation beam at the frequency of the spontaneous SR emission. This is referred to as Stimulated Superradiance (St-SR) [1] [3] [4]. Furthermore, in the case of continuously bunched beam it is possible to extract more energy from the beam by tapering the magnetic undulator period or amplitude after the bunched beam gets trapped and saturates, just as it is routinely done in conventional FEL [5]. This process is termed in the case of a pre-bunched beam: "Tapering Enhanced Stimulated Superradiance Amplification- TESSA". Record high transfer of energy between a periodically bunched electron beam and the radiation field was demonstrated experimentally by this process at optical frequencies by the UCLA FEL group [5] [6].

An interesting question arises whether a bunched beam, emitting SR radiation may interact at saturation and get trapped by its own radiation, and whether in this case it is also possible to enhance the extraction of energy from the beam by undulator tapering. This is therefore a process of Tapering-Enhanced Superradiance (TES). This process raises curiosity, because it is related to the fundamental question of Abraham-Lorentz self-interaction of a single free electron emitting radiation in free space, where the light, faster from the electron, necessarily flies away from the emitting electron. The energy loss of the single electron is explained in classical physics indirectly (through conservation of energy) by the concept of radiation reaction force [7] [8] [9] [10].

Self-interaction of an electron beam with its spontaneously emitted SR radiation is possible in free space with a continuous periodically bunched electron beam (case 1 above), including saturation and self-trapping of the bunched electrons in their self-emitted SR radiation. Also, further energy extraction from the trapped modulated beam by the TES process is possible [1]. Yet, this case is not exactly a demonstration of self-interaction because the radiation emitted in each wiggling period of the undulator slips ahead of the micro-bunch that generated it and interacts only with electrons in microbunches ahead. A more interesting case would be the case of SR from a single short bunch (case 2 above), in which the electrons in the bunch interact and saturate by the radiation they generated (self-interaction). Can TES be realizable in this case? The problem in this case is that in free space the group velocity of the radiation ($c$) is always faster than the velocity of the beam ($v_z$). Therefore, even if phase synchronism is maintained between the electrons and the ponderomotive wave, the bunch slips back after each wiggling period and cannot get trapped in the radiation wavepacket it creates.

A solution to this problem was suggested in [11], [3] and was demonstrated experimentally in the THz frequency regime [12]. The solution is based on performing the interaction within a waveguide that guides a radiation mode along the undulator. The dispersion relation of a waveguide mode enables propagation at slow group velocity $v_{gr} < c$ , and thus it is possible to find an interaction point in which the electron bunch can maintain both group and phase synchronism with ponderomotive wave in which it is trapped (zero- slippage condition).

The analysis of SR and TES of a short bunch at zero-slippage condition cannot be rigorously carried out in the frequency regime because the emission spectrum in this case is very wide and time domain analysis is required to propagate the electrons dynamics. Several works on SR emission below saturation level were carried out using analytical approximations [11] [3] and using a numerical computations code – Wide Bandwidth – that was developed by [13] [14]. A computer code FEL-GPT, that was developed recently by the UCLA group and Pulsar Inc. [15] makes it possible to solve this problem numerically in a general case, including the saturation regime including general beam dynamics and space-charge effects. This code makes it possible to calculate the SR interaction in time domain by presenting the radiation field in a waveguide mode in terms of many longitudinal modes that discretize the modes in the frequency domain and makes it possible to propagate the beam in terms of the spatial variable (z) instead of time (t).

Beyond theoretical analysis, the processes of single bunch SR and TES were demonstrated recently experimentally in the THz frequency regime (at frequencies 0.5-0.7 THz) in a configuration that consists of a tapered helical undulator and a circular cylindrical waveguide [16] [17]. Another experimental project based on a planar tapered undulator and a rectangular waveguide [18] [19] is under way in Israel based on the ORGAD hybrid RF LINAC gun [20] [21].

In this paper we revisit the previous formulations of single bunch SR and TES with an emphasis on developing analytical expressions of the interaction in time domain. This helps to clarify several theoretical subtleties in earlier works, and derive scaling laws of system parameters that are useful for optimization of radiation emission of a TES THz radiation source. Foremost, the transparent analytical formulation in the ideal beam case and its comparison to numerical simulation results, helps to understand the physical processes of the beam-wave interaction in the two-stage process of the unconventional interaction scenario of TES at zero-slippage condition.

For simplifying the analysis of the TES process, we model the interaction region to be composed of two consecutive undulator sections (see Fig. 1). The first section is a uniform period and amplitude magnetic undulator section and the second section is a tapered period or an amplitude tapered undulator (here we analyze only the latter case). For the sake of concreteness, we apply the theory to a specific configuration of the Israeli THz TES experiment [18] [19]. We assume that SR radiation is emitted along the entire two-part undulator. Enhanced tapering-induced radiation extraction from the beam takes place in the second section where the field of the wavepacket signal, built up in the first section, is intense enough to produce a deep enough trap that can keep the electron bunch trapped along the tapered section [22]. The beam acceleration energy is set to satisfy the zero-slippage condition in order to avoid de-trapping of the electrons due to slippage.

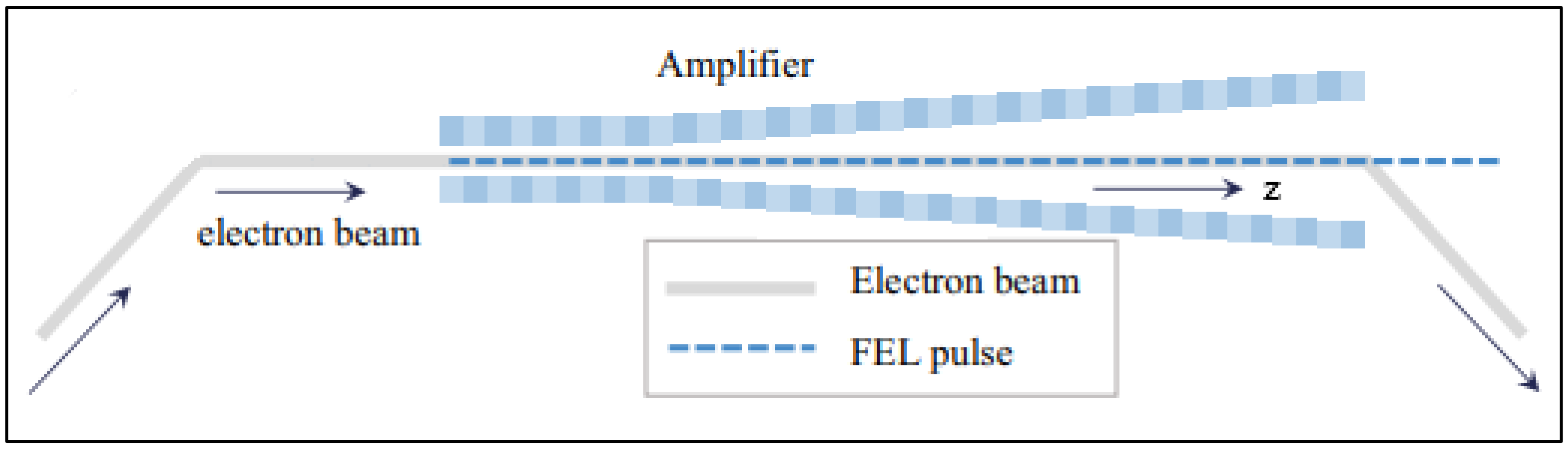

*Fig. 1: Tapering-Enhanced Superradiance (TES) setup composed of a uniform and amplitude-tapered undulator sections.*

## 2. Synchronism and zero-slippage conditions

When undulator synchrotron radiation is generated in a waveguide, its frequency is determined by the dispersion relation of the waveguide radiation mode $k_z(\omega)$ and the synchronism condition:

$$\frac{\omega}{v_z} = k_z(\omega) + k_w \tag{2}$$

$$k_z(\omega) = \frac{\sqrt{\omega^2 - \omega_{co}^2}}{c} \tag{3}$$

Where $\omega = 2\pi f$ is the radiation frequency, $v_z$ is the average axial velocity of the undulating electron beam, $k_w$ is the undulator wavenumber, $\omega_{co}$ is the cut-off frequency of the mode.

These two equations have in general two solutions (see Fig. 2):

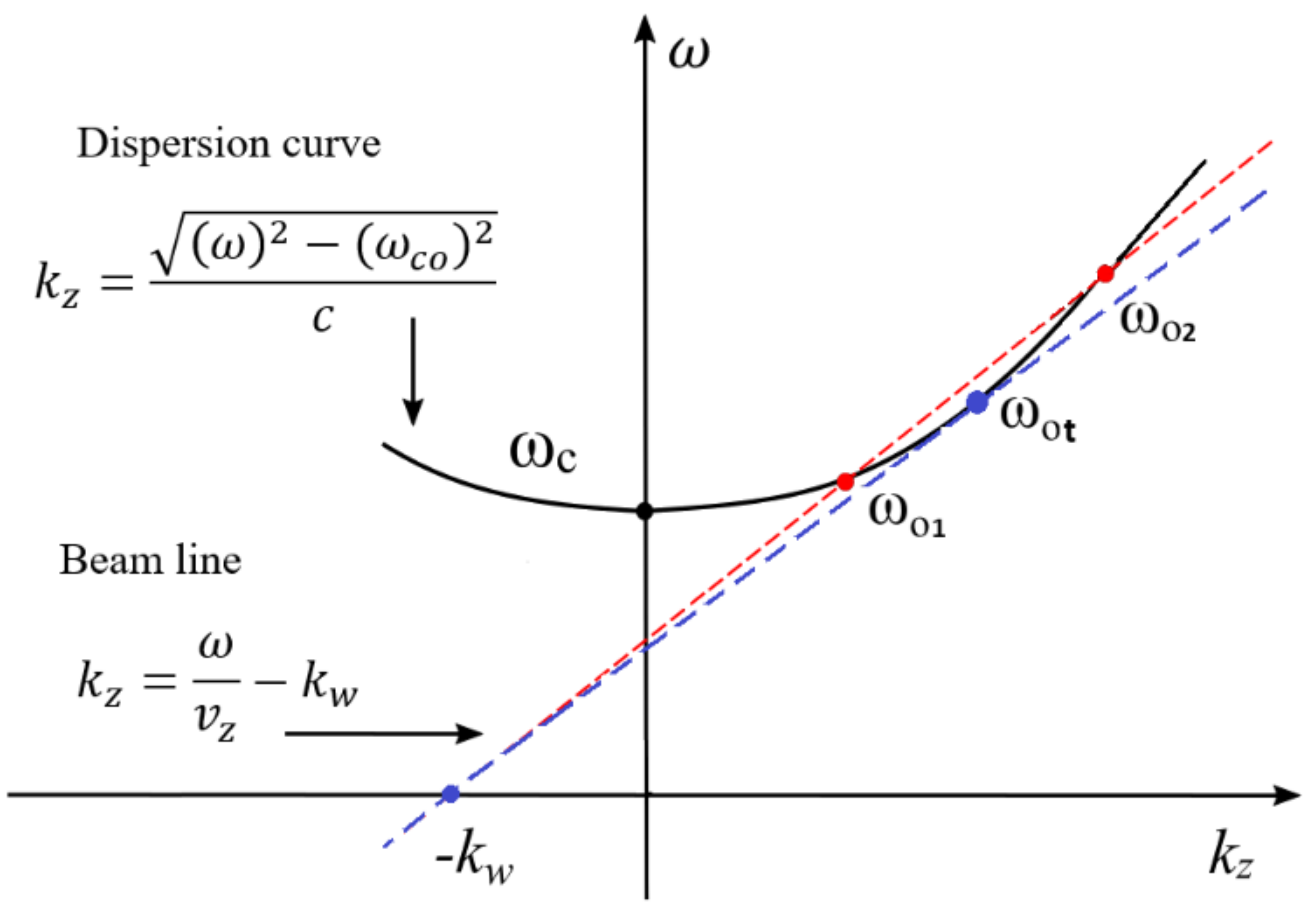

*Figure 2: Intersection of the beam line and the dispersion curve of a waveguide mode. At normal FEL operation there are two solutions to the phase synchronism condition (Eq. 2) (red line). At zero-slippage condition (blue line) the curves are tangent, and the group velocity of the radiation mode is equal to the axial velocity of the beam $v_z$ .*

$$\omega_{0,12} = \gamma_z^2 \beta_z c k_w \left( 1 \pm \sqrt{\left( \beta_z^2 - \left( \frac{\omega_{co}}{\gamma_z k_w c} \right)^2 \right)} \right) \tag{4}$$

Where $\gamma_z = (1 - \beta_z^2)^{-1/2}$ . In the case of a planar undulator $\gamma_z = \sqrt{\frac{\gamma^2}{1+(\frac{K^2}{2})}}$ where $K = \frac{eB_w}{m_e c k_w}$ is the undulator parameter

At the upper frequency solution $\omega_{02}$ the group velocity of the radiation mode (the slope of the dispersion curve – see Fig. 2) is higher than the electron axial velocity, which means that the radiation wavepacket emitted by the electron, slips ahead of the emitting electron. At the lower frequency solution, the group velocity of the radiation is smaller than the speed of the electron and the emitted radiation wavepacket lags the electron. At the particular beam velocity $\beta_z = \left( \frac{\omega_{co}}{\gamma_z k_w c} \right)$, the two solutions coincide

$$\omega_{0,t} = \gamma_z^2 \beta_z c k_w \tag{5}$$

and the beamline is tangent to the mode dispersion curve (Fig. 1). At this condition, the radiation wavepacket, emitted spontaneously by the electron, does not slip away from the electron and resides with it as the electron propagates along the undulator.

This special operating point of undulator radiation (spontaneous emission of FEL) is of special interest when the undulator radiation is excited by an ultrashort electron beam that satisfies the superradiance condition (1). In the absence of slippage of the radiation wavepacket generated by the beam, the radiation stays as a short duration signal, keeping in pace with the electron beam. The frequency bandwidth of the coherent signal is very wide because the phase synchronism condition (1) is maintained approximately for a wide range of frequencies around the exact center frequency (5).

This combination of superradiance and zero-slippage condition is of special interest at TeraHertz frequencies, where Eq. 1 can be easily satisfied with common sub-picoSecond electron bunches available in photo-cathode RF-Gun injectors. This can be used to generate ultra-short intense THz radiation pulses in a waveguide FEL configuration with a uniform wiggler without stimulated emission. As discussed above, the extraction of energy from the short bunch can be enhanced beyond the saturation point if the bunch gets trapped in the ponderomotive potential and the undulator is tapered, while keeping the zero-slippage condition all along to prevent beam de-trapping.

### 3. Radiation field excitation in the frequency and time domains.

To solve the problem of radiation emission from e-beam radiation devices we use a formulation of modal expansion of the excited radiation field in the frequency domain [1]. The space-dependent radiation field is expanded in terms of the eigenmodes $\tilde{\boldsymbol{\epsilon}}_{\boldsymbol{q}}(r_\perp)e^{ik_{qz}z}$ of the waveguide, which is assumed to be a complete set of orthogonal modes propagating in the z direction:

$$\breve{\mathrm{E}}(r,\omega)=\sum_{\pm q}\breve{\mathrm{C}}_q\,(z,\omega)\tilde{\boldsymbol{\epsilon}}_{\boldsymbol{q}}(r_\perp)e^{ik_{qz}z} \tag{6}$$

$C_q(z,\omega)$ is the slowly growing field amplitude of radiation mode q at spectral frequency $\omega$ when interacting with an electron beam along its propagation path (z). The formulation is general for any waveguide, here we consider a rectangular waveguide and a planar magnetic undulator polarized in the y direction.

The evolution of the mode amplitudes is derived from Maxwell equations in the frequency domain

$$\frac{dC_q(z,\omega)}{dz}=\frac{-1}{4P_q}\int \boldsymbol{J}(r,\omega)\cdot\tilde{\boldsymbol{\epsilon}}_{\boldsymbol{q}}^{\,*}(r_\perp)e^{-ik_{qz}z}dA \tag{7}$$

where $P_q$ is an arbitrary power normalization parameter of mode q.

In previous publications [1]the frequency dependent parameters in (6), (7) were interpreted as Fourier transforms of the real-time currents and field: $J(r,\omega)=\int_{-\infty}^{\infty}\boldsymbol{J}(r,t)\exp(i\omega t)\,dt$, $E(r,\omega)=\int_{-\infty}^{\infty}E(r,t)\exp(i\omega t)\,dt$. However, in the mode excitation model (6) (7) it is assumed that the electron beam currents interact only with forward going radiation modes $\tilde{\mathcal{E}}_q(r_\perp)e^{-ik_{qz}z}$ ($k_{qz}>0$). This leaves negative frequencies undefined. A more rigorous formulation was presented by Pinhasi and Lurie [14] [13]. Starting with a real field in time domain $E(r,t)$, they introduce analytical extension of the time domain field to the complex field domain by the use of Hilbert transformation:

$$\breve{E}(r,t) = E(r,t) - i\int_{-\infty}^{\infty}\frac{E(r,t')}{t-t'}dt' \tag{8}$$

With this definition, a phasor-like function of the field can be defined in terms of the positive frequencies part of the Fourier transform of the analytically extended field $E(r,\omega)$:

$$\breve{E}(r,\omega) = 2E(r,\omega)u(\omega) \tag{9}$$

where $u(\omega)$ is the step function and the quasi-phasor field $\breve{E}(r,\omega)$ is the solution of the excitation equations (6), (7).

Correspondingly, the time domain field can be retrieved from the phasor field by the transformation:

$$E(r,t) = \frac{1}{2\pi}\int_{-\infty}^{\infty}E(r,\omega)e^{-i\omega t}d\omega = \Re\left\{\frac{1}{2\pi}\int_{0}^{\infty}\breve{E}(r,\omega)e^{-i\omega t}d\omega\right\} \tag{10}$$

The spectral power density of the radiation mode is then calculated using Parseval theorem and given in terms of positive frequencies $\omega > 0$ :

$$\frac{dW_q}{d\omega} = \frac{1}{2\pi}P_q\left|\breve{C}_q(z,\omega)\right|^2 \tag{11}$$

We note that the mathematical subtlety of using quasi-phasor formulation instead of Fourier transformation is not expected to be significant when the radiation spectrum is narrow, however, we expect the spectrum of SR at zero-slippage condition to be very wide relative to the center frequency. In the following we will use the quasi-phasor formulation and find out the limits of the Fourier transform approximation.

For a particulate current:

$$\boldsymbol{J}(\boldsymbol{r},t) = \sum_{j=1}^{N} -e\mathbf{v_j}(t)\delta\left(\boldsymbol{r}-\boldsymbol{r_j}(t)\right) = \sum_{j=1}^{N} -e\frac{\boldsymbol{v_j}(t)}{v_{zj}}\delta\left(\boldsymbol{r}_\perp - \boldsymbol{r}_{\perp\boldsymbol{j}}(t)\right)\delta\left(t - t_j(z)\right) \tag{12}$$

N is the number of electrons, e is the electron charge, $\boldsymbol{v}_j$is the jth electron velocity vector. Also here we need to present the current in frequency domain in terms of quasi-phasor notation:

$$\breve{\boldsymbol{J}}(\boldsymbol{r},\omega) = 2u(\omega)\int_{-\infty}^{\infty}\boldsymbol{J}(r,t)e^{i\omega t}dt = 2u(\omega)\sum_{j=1}^{N} -e\boldsymbol{v_j}(t_j(z))\delta\left(\boldsymbol{r}_\perp - r_{\perp j}\left(t_j(z)\right)\right)e^{i\omega t_j(z)} \tag{13}$$

Substitution of (13) in (7) in quasi-phasor notation, results in:

$$\frac{d\breve{C}_q(z,\omega)}{dz} = \frac{1}{2P_q}\sum_{j=1}^{N} e\frac{\boldsymbol{v_{j\perp}}(z)}{v_{zj}}\cdot\tilde{\boldsymbol{\epsilon}}_{\boldsymbol{q}}^{\,*}\left(r_{\perp j}\right)e^{i\omega t_j(z) - ik_{qz}z} \tag{14}$$

where $t_j(z) = t_{0j} + z/v_{zj}$.

Now we assume that all electrons in the bunch follow the same trajectories $\boldsymbol{r}_{\perp e}(z)$ except that they enter at different entrance times $t_{0j}$. We further assume that the energy change of the beam is negligible (no saturation effect), and $v_{zj} = v_{z0}$ for all electrons, then Eq. 14 can be integrated directly along the interaction length. Assuming $C_q(0) = 0$ (spontaneous emission),

$$\check{C}_q(L_w,\omega) = \frac{1}{2P_q}\Delta\breve{W}_{qe}\sum_{j=1}^{N} e^{i\omega t_{0j}} \tag{15}$$

where

$$\Delta\breve{W}_{qe} = -\frac{e}{v_{z0}}\int_0^{L_w} \boldsymbol{v}_{\perp\boldsymbol{e}}(z)\cdot\tilde{\boldsymbol{\epsilon}}_{\boldsymbol{q}}^{\,*}(r_{\perp e})e^{i\left(\frac{\omega}{v_{z0}}-k_{zq}\right)}dz \tag{16}$$

This can be interpreted as a spectral work function of the radiation wavepacket emitted spontaneously by a single electron. Eq. 15 is interpreted as the frequency representation of the interference of all wavepackets emitted by the beam. It can be written as:

$$\check{C}_q(L_w,\omega) = \frac{1}{2P_q}\Delta\breve{W}_{qe}NM_b e^{i\omega t_0} \tag{17}$$

$t_0$ is the entrance time of the bunch.

$$M_b = \frac{1}{N}\sum_{j=1}^{N} e^{i\omega(t_{0j}-t_0)} \tag{18}$$

$M_b$ is the bunching factor of the beam. For a Gaussian distribution of the bunch f(t)=$\frac{1}{\sigma\sqrt{2\pi}}e^{-\frac{1}{2}\left(\frac{t}{\sigma}\right)^2}$

$$M_b = e^{-\frac{(\omega\sigma_t)^2}{2}} \tag{19}$$

and clearly, in order to keep the bunching parameter near 1, the superradiance condition (1) must be satisfied.

For the case of a planar undulator $\boldsymbol{v}_{\perp\boldsymbol{e}}(z) = Re\{\tilde{\boldsymbol{v}}_w e^{-ik_w z}\} = \boldsymbol{v}_w\cos(k_w z)$, and the wavepacket spectral work function (16) is [ 1 ] (see Appendix A):

$$\Delta\breve{W}_{qe} = -e\frac{\boldsymbol{v}_{w0}\cdot\tilde{\boldsymbol{\epsilon}}_{\boldsymbol{q}}^{\,*}(\boldsymbol{r}_{\perp\boldsymbol{0}})}{2v_{z0}}L_w sinc\left(\frac{\theta L_w}{2}\right)e^{\frac{i\theta L_w}{2}} \tag{20}$$

where $v_w = \frac{cK}{\gamma}$ is the transverse velocity amplitude of the electron inside the wiggler.

$L_w = \lambda_w\cdot N_w$ is the interaction length.
The detuning parameter is:

$$\theta(\omega) = \frac{\omega}{v_z} - k_{zq}(\omega) - k_w \tag{21}$$

where $v_z$ is the axial velocity of the e-beam inside the wiggler, $k_{zq}(\omega)$ is the dispersion relation (3), $k_w$ is the undulator wavenumber.

Considering interaction with a single radiation mode, and using (17), (20) (Append. A), the quasi-phasor field (6) is given by:

$$\boldsymbol{E}(r_\perp, Lw, \omega) = A\frac{\tilde{\boldsymbol{\epsilon}}_{\boldsymbol{q}}(\boldsymbol{r}_\perp)}{\left|\tilde{\boldsymbol{\epsilon}}_{\boldsymbol{q}}(\boldsymbol{r}_{\perp\boldsymbol{0}})\right|}F_q(\omega) \tag{22}$$

$$F_q(\omega) = sinc\left(\frac{\theta(\omega)\mathrm{z}}{2}\right)e^{\frac{i\theta(\omega)\mathrm{z}}{2}}e^{ik_z(\omega)\mathrm{z}}e^{i\omega t_0} \tag{23}$$

$$A = N\frac{M_b(\omega_0)Z_q}{2A_{em}}\frac{eK}{\beta_z\gamma}L_w \tag{24}$$

The effective area of the mode is defined via the mode power normalization parameter and the field of the mode at the transverse position of the electron path axis $\boldsymbol{r}_{\perp 0}$

$$P_q = \frac{\left|\tilde{\boldsymbol{\epsilon}}_{\boldsymbol{q}}(r_{\perp 0})\right)|^2}{2Z_q}A_{em,q} \tag{25}$$

Here $Z_q$ is the mode impedance.

Substituting (22) in (10), the real time-dependent field is:

$$E(r,t) = \frac{A}{2\pi}\Re\left\{\int_0^\infty e^{-i\omega(t-t_0)}e^{\frac{i\theta L_w}{2}}sinc\left(\frac{\theta L_w}{2}\right)e^{ik_zL_w}d\omega\right\} \tag{26}$$

We now calculate the spectral energy (11). Substituting (17), ((20) into (11) (Append. A), the spectral energy of SR is [1]:

$$\frac{dW_q}{d\omega} = \frac{e^2}{16\pi}\frac{N^2M_b^2Z_q|\tilde{a}_w|^2}{(\beta_z\gamma)^2}\frac{L_w^2}{A_{emq}}sinc^2\left(\frac{\theta(\omega)L_w}{2}\right) \tag{27}$$

$$(|\tilde{a}_w| = K)$$

The total radiative energy of the SR pulse is:

$$Wq = \int_0^\infty \frac{dW_q}{d\omega}d\omega = \frac{e^2}{16\pi}\frac{N^2M_b^2Z_qK^2}{(\beta_z\gamma)^2}\frac{L_w^2}{A_{emq}}\Delta\,\omega \tag{28}$$

where we define the spectral bandwidth:

$$\Delta\omega = \int_0^\infty sinc^2\left[\frac{\theta(\omega)L_w}{2}\right]d\omega \tag{29}$$

**4. Superradiance at zero-slippage condition**

For the zero-slippage case (after quadratic expansion – see Appendix B):

$$\theta(\omega) = -\frac{1}{2}D(\omega-\omega_0)^2 \tag{30}$$

$D = \omega_{co}^2/c(\omega_0\beta_z)^3$

We first evaluate the spectral energy and the spectral bandwidth of SR at zero-slippage condition. Define:

$$\tau = \frac{\sqrt{DL_w}}{2} \tag{32}$$

$$\Delta\omega = \int_0^\infty sinc^2[\tau^2(\omega-\omega_0)^2]d\omega = \frac{1}{\tau}\int_{-\omega_0\tau}^\infty sinc^2u^2du \tag{33}$$

$$(u = \tau(\omega-\omega_0))$$

The quadratic expansion is valid for $\omega_0\tau > \sqrt{2\pi}$. We have found that in this range one can replace the lower limit of the integral $-\omega_0\tau$ with $-\infty$ and to a good approximation we can use the formula $\int_{-\infty}^{\infty} sinc^2u^2du = \frac{4}{3}\sqrt{\pi} = 2.36$. Therefore, within this approximation, the bandwidth is fully specified by the parameter $\tau$:

$$\Delta\omega = \frac{2.36}{\tau} \tag{34}$$

We present the spectral energy function $\left|F_q(\omega)\right|^2 = sinc^2(\frac{\theta(\omega)L_w}{2})$ in Fig. 3. To check the validity of the quadratic expansion with $\theta(\omega)$ given by Eq. 21 we also show (overlayed) the spectral energy function without the quadratic expansion (30). This is shown for $N_W$=5 and $N_W$=20. The bandwidth expression Eq. (34) is shown and agrees well with the FWHM bandwidth value of the curve.

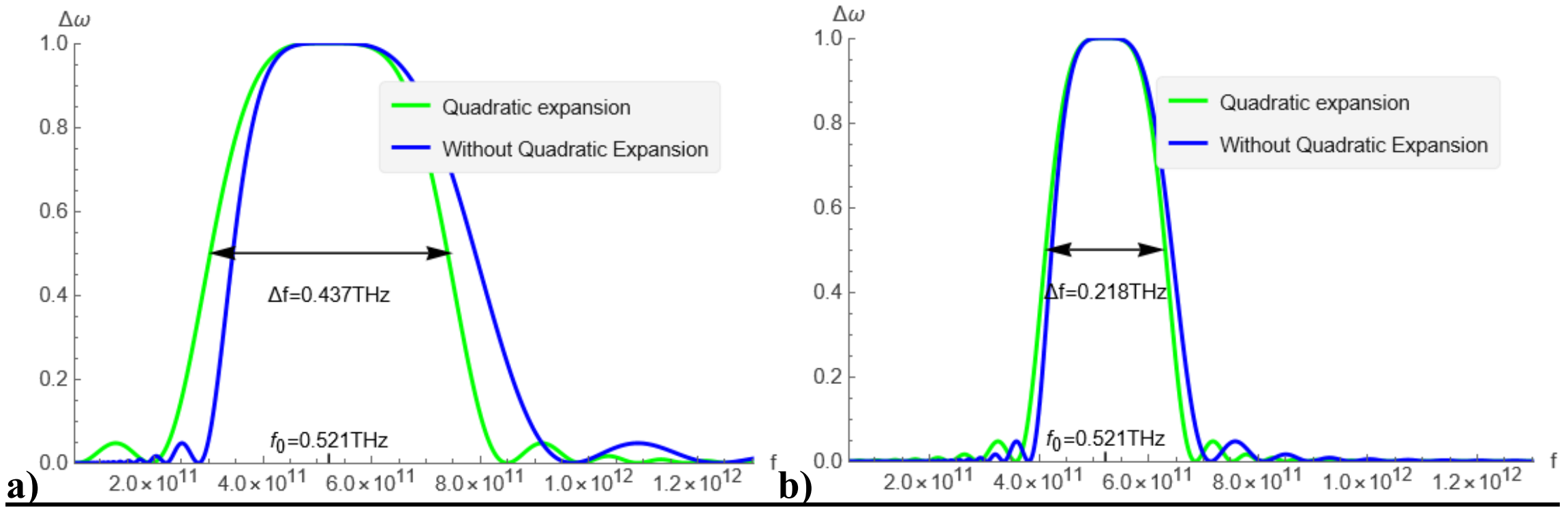

*Fig. 3: The spectral energy function $sinc^2(\frac{\theta(\omega)L_w}{2})$ at zero-slippage conditions with and without quadratic expansion (overlaid) for (a) $N_W$=5 and (b) $N_W$=20*

There is quite a good agreement between the curves with and without quadratic expansion and the bandwidth expression (34) proves valid. This holds even in the wide frequency band case of $N_W$=5, for which (32) - $\tau = 0.86\ pSec$ , and $\Delta f = \frac{\Delta\omega}{2\pi} = 0.437\ THz$ (34), which is a significant fraction of the center frequency $f_0 = 0.521\ THz$. This is quite a good agreement, indicating that the quadratic approximation is valid here. These parameters were calculated for the parameters of Table 1,

Table 1: Parameters of the TES setup

| | Value |
|---|---|
| Beam Energy E | 6Mev |
| Charge Q | -50pC |
| Lambda wiggler | 0.056 m |
| Peak magnetic field | 0.2195 T |
| Undulator parameter K | 1.147 |
| Waveguide height b | 2.8448mm |
| Waveguide width a | 5.689mm |
| Synchronism frequency | 0.521THz |
| Tapering rate dK/dz | -0.1847 [1/m]<br>( $\psi_r = 45°$<br>(for E0=13Mv/m) |

Using this approximation, the total SR radiative energy can be presented in terms of an analytic expression:

$$Wq = \frac{e^2}{16\pi}\frac{N^2 M_b^2 Z_q K^2}{(\beta_z\gamma)^2}\frac{L_w^2}{A_{emq}}\frac{2.36}{\tau} \qquad (35)$$

Because the bandwidth gets narrower the longer is the undulator (32, 33), and scales like $N_w^{-1/2}$, the SR energy in the case of zero-slippage scales like $N_w^{3/2}$ as shown in Fig. 4. It also scales with the bunch charge as $Q^2$. In Fig. 4 we show the scaling with $N_w$ for a uniform undulator for the parameters of Table 1 and Q=50pC

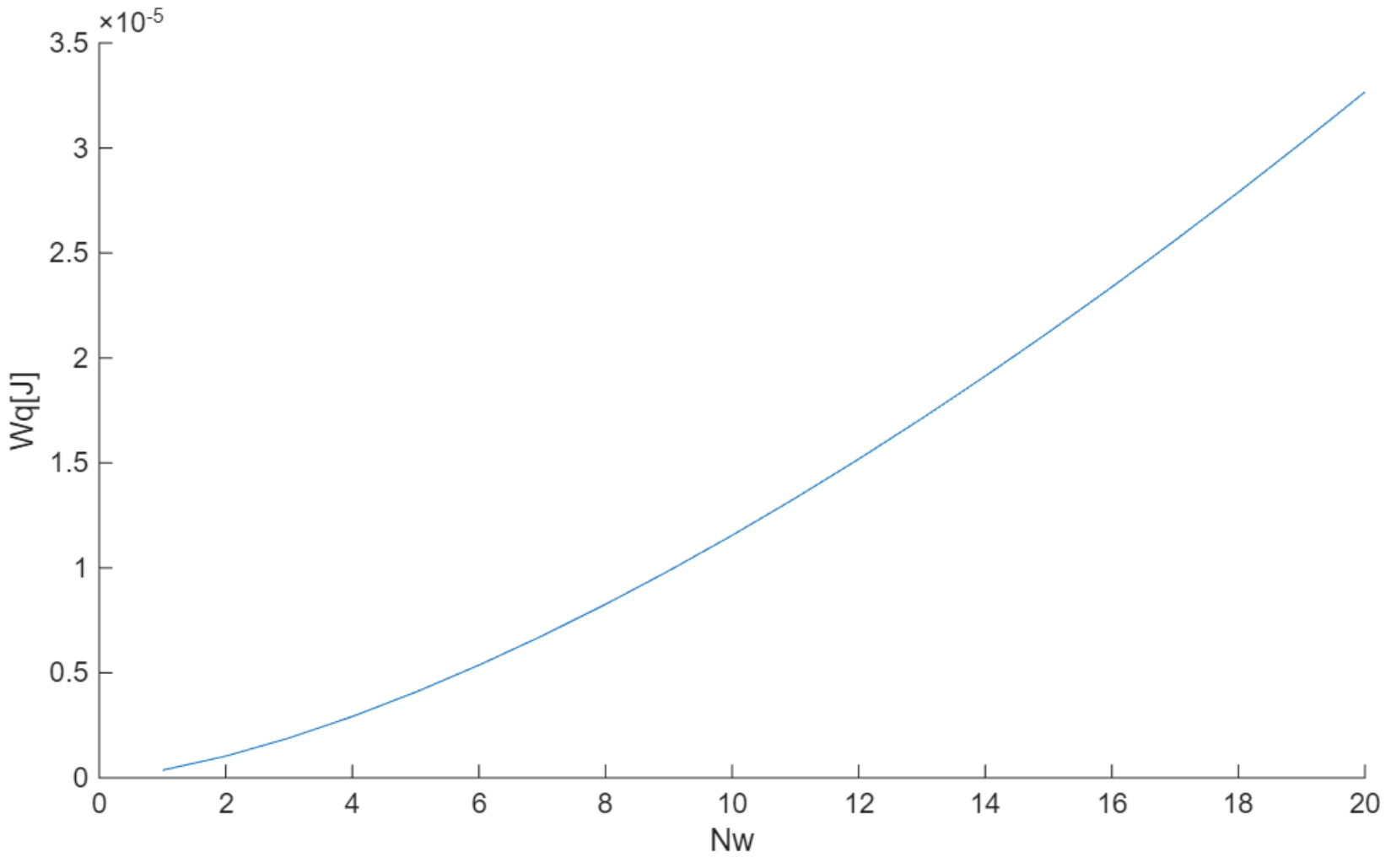

*Figure 4: Plot of the analytical expression (35) of SR radiated energy as function of $N_w$ for a uniform undulator (N_w=20) and Q=50pC, showing scaling as $N_w^{3/2}$ .*

**5. Field amplitude in time domain of SR at zero-slippage condition**

We use (10) to transform the frequency domain expression of the field (22) to the time domain:

$$E(r,t) = A\frac{\tilde{\mathcal{E}}_q(\boldsymbol{r}_\perp)}{|\tilde{\mathcal{E}}_{\boldsymbol{q}}(\boldsymbol{r}_{\perp\boldsymbol{0}})|}f_q(t) \tag{36}$$

$$f_q(t) = \Re\left\{\int_0^\infty F_q(\omega)e^{-i\omega t}\,d\omega\right\} \tag{37}$$

and $f_q(t)$, the field waveform function, is (using (23)):

$$f_q(t) = \Re\left\{\frac{1}{2\pi}\int_0^\infty e^{-i\omega(t-t_0)}\,e^{\frac{i\theta L_w}{2}}sinc\left(\frac{\theta L_w}{2}\right)e^{ik_zL_\omega}d\omega\right\} \tag{38}$$

It can be presented relative to the arrival time of the bunch to the end of the undulator: $t' = t - t_0 - L_w/v_{z0}$:

$$f_q\left(t - t_0 - \frac{L_w}{v_z}\right) = \Re\left\{\frac{1}{2\pi}\int_0^\infty e^{-i\omega\left(t-t_0-\frac{L_w}{v_z}\right)}\,e^{\frac{i\theta L_w}{2}}sinc\left(\frac{\theta L_w}{2}\right)e^{ik_zL_\omega}d\omega\right\} \tag{39}$$

In the case of zero-slippage, we show in Append. C that after substitution of the quadratic expansion of the mode wavenumber (B1) (B3), the field waveform function can be written in terms of a dimensionless complex function $g(t'/\tau)$

$$f_q(t') = \frac{1}{\tau}Re\left\{e^{-i\omega_0 t'}g\left(\frac{t'}{\tau}\right)\right\} \tag{40}$$

$$g\left(\frac{t'}{\tau}\right) = \frac{1}{2\pi}\int_{-\omega_0\tau}^{\infty} e^{-iu\frac{t'}{\tau}}sinc(u^2)e^{-iu^2}du \tag{41}$$

The function $g(t'/\tau)$ depends only on the parameter $\omega_0\tau$ in the lower limit of the integral. Considering (34), $\omega_0\tau = 2.36\omega_0/\Delta\omega$. For narrow enough relative bandwidth of the radiation $\Delta\omega/\omega_0$, the integration can be carried out from $-\infty$ to $\infty$ and this function can be presented in terms of a universal function:

$$g_u\left(\frac{t'}{\tau}\right) = \frac{1}{2\pi}\int_{-\infty}^{\infty} e^{-iu\frac{t'}{\tau}}sinc(u^2)e^{-iu^2}du \tag{42}$$

Then the time dependence of the field on the beam axis upon exiting the undulator can be expressed in terms of this complex universal function as:

$$E(r_{\perp 0}, L_w, t) = Af_q(t') = \frac{A}{\tau}\left|g_u\left(\frac{t'}{\tau}\right)\right| \cos\left(\omega_0 t' - \phi_g\left(\frac{t'}{\tau}\right)\right) \quad (43)$$

where $|g_u(t'/\tau)|$ and $\phi_g(t'/\tau)$ are the amplitude and phase of $g_u(t'/\tau)$ respectively. These are displayed in Fig. 5a and b respectively. In Fig. 5 we also overlay the curve of equation (41), calculated for $\omega_0\tau = 2.36$, corresponding to the case of $N_w = 5$. It shows that the universal complex function (42) is a good approximation even in this wide bandwidth case (Fig. 3a).

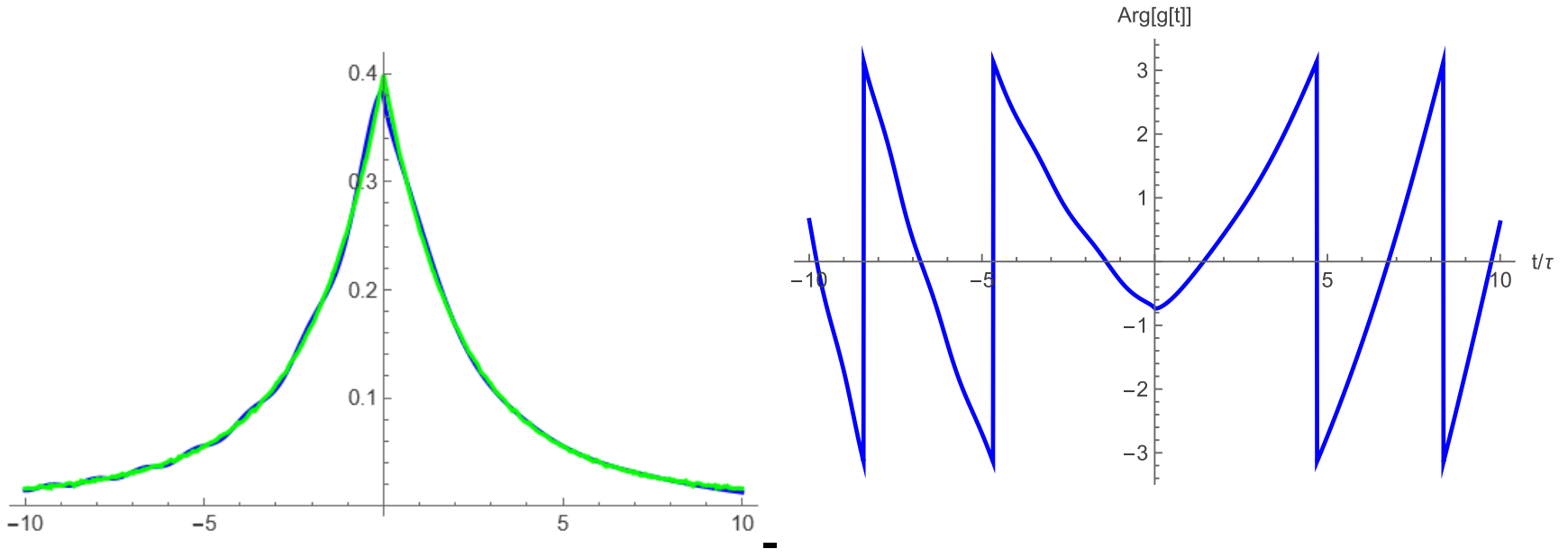

*Figure 5: Plot of the universal waveform complex function $g_u(t'/\tau)$ (eq. 42). left:* $|g(t'/\tau)|$ (Eq. 42 in green), overlayed: $g(t'/\tau)$ (Eq.41 in blue) right: phase of $g_u(t'/\tau)$.

From (41) and Fig. 5 we get a compact analytic expression for the field amplitude at the end of the uniform wiggler section. With $|g_m| = 0.37$, derived from Fig. 5:

$$E_{max} = \frac{A}{\tau}|g_m| = 0.37\frac{A}{\tau} \quad (44)$$

This parameter is needed for the calculation of the trap depth in the trapping limit and in the entrance to the tapered undulator section in the following section. For the parameters of Table 1 we find out (24 ) for Q=-50 pC and $N_w = 5$ that A=3x10$^{-5}$ V-Sec/m, $\tau = 8.6x10^{-13}Sec$, and consequently $E_{max} = 13\ MV/m$

For the same parameters we calculate the entire time dependence of the field waveform (41, 43) at the end of the uniform undulator section $N_w = 5$ displayed in Fig. 6. Eq. 43 and Fig 6 show that the field of the SR wavepacket looks like a carrier wave of frequency $\omega_0$ modulated by an envelope amplitude $|g_u(t'/\tau)|$ and chirping phase $\phi_{gu}(t'/\tau)$ that are plotted in Fig 5. The curve in Fig. 6 shows $E_{max} = 13\ MV/m$, in good agreement with the calculation from (44).

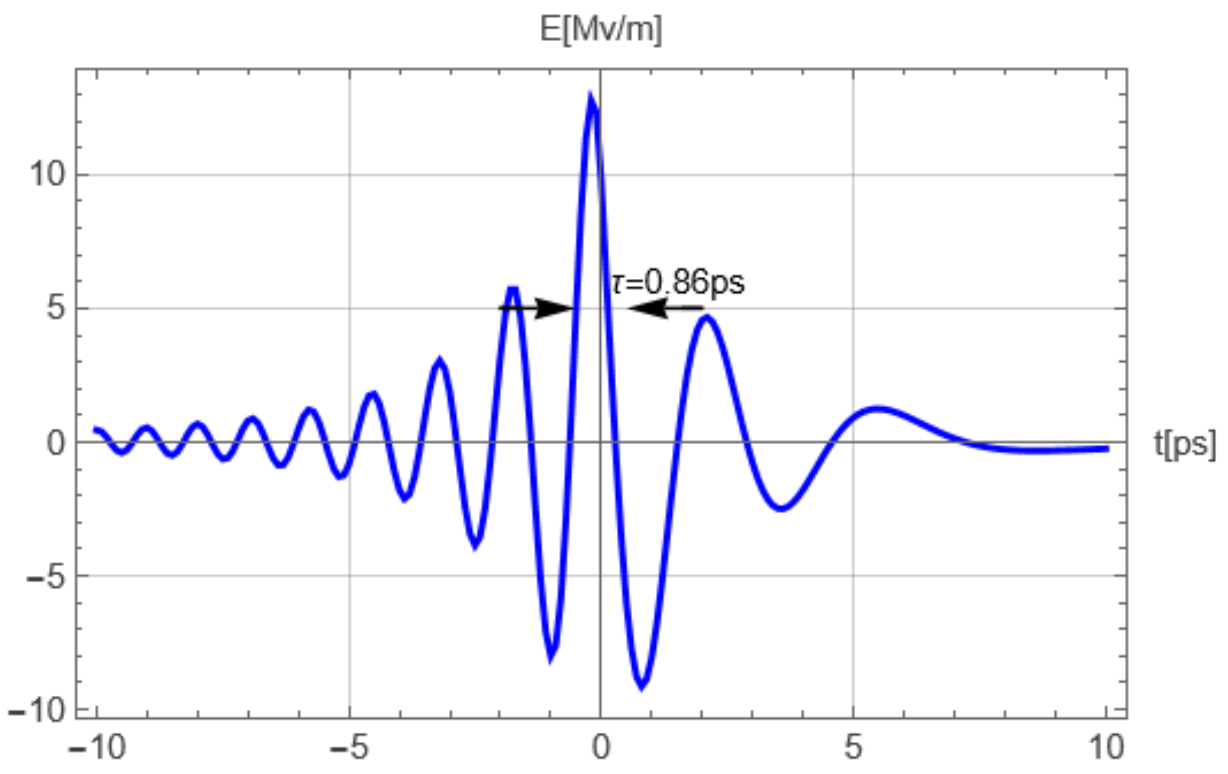

Fig. 6 *The field waveform in time domain at the end of the uniform section $N_w$=5. The calculated wavepacket duration (Eq. 32) is shown*

### 6. Energy extraction in the second (TESSA) wiggler section

As in tapered FEL [22] and TESSA theory [23], the derivation of the radiative energy extraction is indirect, and is based on calculation of the energy loss by the electron beam and on conservation of energy. In this section we assume that an ideal electron beam bunch is fully trapped and loses energy in favor of the radiation mode due to the tapering of the undulator magnetic field amplitude.

Neglecting the inner dynamics of the electrons in the trap and neglecting de-trapping, the energy drop of the resonant (deeply trapped) electron is [23] [18]:

$$\frac{d\gamma_r^2}{dz} = -k_0 K\, K_L \sin\psi_r \tag{45}$$

where $K = \frac{eB_w}{mc\, k_w}$, $k_0 = \omega_0/c$,

$$K_L = e\frac{E_m}{k_0 mc^2} \tag{46}$$

and $\psi_r$ is the bunch phase relative to the ponderomotive potential wave. For a linear magnetic field tapering rate $\frac{dK}{dz}$ , it is defined through the relation:

$$sin\,\psi_r = \frac{dK}{dz}/k_w K_L \tag{47}$$

In the special case of zero-slippage condition, the synchronism condition (5) is (in the highly relativistic limit):

$$\gamma_{zr}^2 = \frac{k_0}{k_w} \approx const \tag{48}$$

Assuming the frequency is constant, this condition must be kept also in the tapered undulator section and from the relation $\gamma_r^2 = \gamma_{zr}^2\left(1+\frac{K^2}{2}\right)$

$$\frac{d\gamma_r^2}{dz} = \gamma_{zr}^2 K\frac{dK}{dz} \tag{49}$$

Substituting in (45) and using the synchronism condition (48)

$$\frac{dK}{dz} = -k_w K_L\, sin\,\psi_r \tag{50}$$

This sets a limit on the tapering rate to keep $sin\,\psi_r < 1$:

$$\left|\frac{dK}{dz}\right| < k_w K_L \tag{51}$$

otherwise the pondermotive potential trap, presented in phase-space by the separatrix (Fig. 7), vanishes. At zero slippage conditions, the trap separatrix is given by [18]:

$$\delta\gamma = \pm\gamma_{z0}\sqrt{K_L KJJ\,[\cos(\psi_r) + \cos(\psi) - (\pi - \psi_r - \psi) sin(\psi_r)]} \quad (52)$$

and the depth of the trap is given by

$$\delta\gamma_{trap} = 2\delta\gamma(\psi_r) = 2\gamma_{z0}\sqrt{2KK_L JJ}\sqrt{\cos\psi_r + \left(\psi_r - \frac{\pi}{2}\right) sin\psi_r} \quad (53)$$

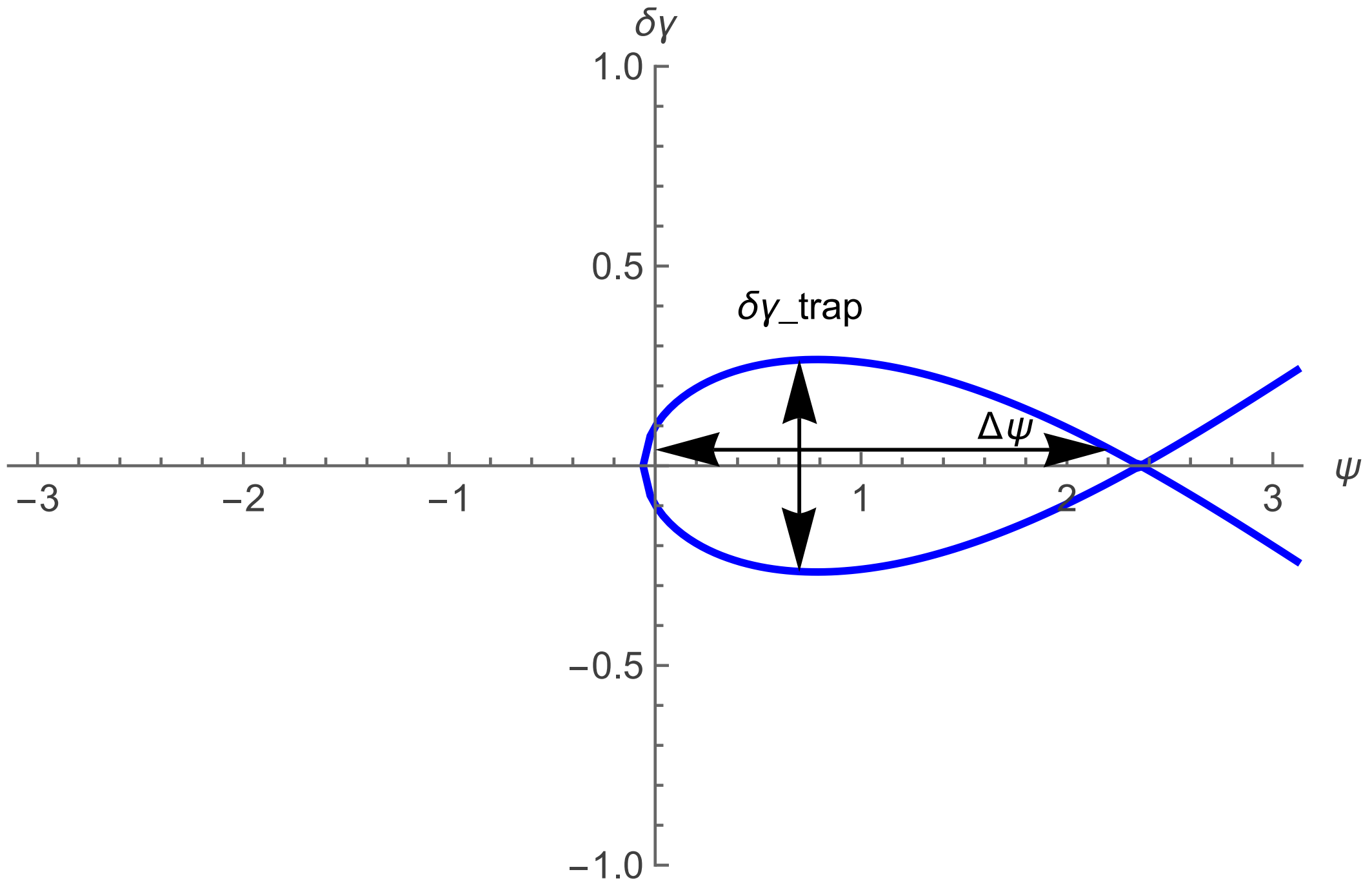

Fig. 7: The separatrix representing the ponderomotive potential trap in a tapered undulator in $\delta\gamma - \psi$ phase-space. It is drawn based on Eq. 52 for the beam parameters of Table1 and $\psi_r = \pi/4$.

This formulation is useful for setting limitations on the tapered section design. Once one evaluates $K_L$ from the radiation field at the entrance to the tapered undulator (46), One can choose a tapering rate $\left|\frac{dK}{dz}\right|$ that satisfies the inequality (51) and evaluate the corresponding resonant phase of the bunch $\psi_r$. Furthermore, based on (52), (53), one can set limits on the energy spread and bunch duration required in order to keep the bunch within the trap:

$$\delta\gamma_{spread} < \delta\gamma_{trap}(\psi_r) \quad (54)$$

$$2\sigma_t < \Delta\psi < 2\pi \quad (55)$$

We now calculate the beam energy drop in the tapered section by integration of Eq. 49. We assume an ideal fully trapped beam and a moderate change in the wiggler strength and the radiation field amplitudes along the wiggler $K(z) \simeq K_z(z_{02})$, $K_L(z) \simeq K_L(z_{02})$, where $z_{02}$ is the entrance point of the second wiggler,

$$\gamma_r^2(L_{w2}) - \gamma_r^2(z_{02}) \simeq -\gamma_{zr}^2 k_w K(z_{02}) K_L(z_{02}) \sin(\psi_r)\, L_{w2}$$

$$\Delta\gamma_r \simeq -\frac{\gamma_{zr}^2}{2\,\gamma_r(z_{02})} k_w K(z_{02}) K_L(z_{02}) \sin(\psi_r)\, L_{w2}$$

With the zero-slippage synchronism condition in the highly relativistic limit, $\gamma_{zr}^2 = \frac{k_0}{k_w}$ (8):

$$\Delta\gamma_r \simeq -\frac{k_0}{2\,\gamma_r(z_{02})} K(z_{02}) K_L(z_{02}) \sin(\psi_r)\, L_{w2} \tag{56}$$

The radiative energy gain in the second section is then:

$$\Delta W_2 = Nmc^2 \Delta\gamma_r = \mathrm{N}mc^2 \frac{k_0}{2\,\gamma_r(z_{02})} K(z_{02}) K_L(z_{02}) \sin(\psi_r)\, \lambda_w N_{w2} \tag{57}$$

## 7. Comparison of the analytical model to numerical GPT simulations.

The analytical model has a limited scope even in the analysis of TES with an ideal beam. It is quite reliable for representing the energy and field growth of the SR radiation wavepacket in the uniform undulator section according to the derivations in sections 4 and 5. However, it cannot represent the saturation process of the beam (synchrotron oscillation) when the bunch gets trapped in the tapered section, or even in a uniform undulator (see the monotonic growth in Fig 4). On the other hand, the analytical model for the beam dynamics in the tapered (TESSA) undulator section (Section 6) is valid in the nonlinear deep-trapping regime and provides a reasonable estimate of the energy extraction of the trapped e-beam bunch. Through the conservation of energy principle, it is used to evaluate the energy transfer to the radiation wavepacket. However, this model cannot produce the field amplitude and time-domain waveform of the emitted radiation. In general, we expect the output radiative energy to be the sum of the SR and tapering (SASE) contributions $\Delta W_{tot} = \Delta W_{SR} + \Delta W_{SASE}$, but it is hard to determine from the analytical models where the trapping takes place, and the TESSA process starts. Therefore, the comparison of the analytical models to the numerical computation is useful only in parts.

The numerical computation model of FEL-GPT has a much wider validity. It includes saturation (Synchrotron oscillation of trapped electrons), space-charge effects and finite beam quality parameters (realistic beam) effects. However, the algorithm of the FEL-GPT code may have some limitations in calculating the radiation field waveform. It is based on expanding the radiation field in terms of many longitudinal radiation modes within the length of the interaction length, such that they can represent the entire bandwidth of the expected radiation emission. The resultant field is the coherent sum of the interference of all the modes at the end of the interaction length. This algorithm may be sensitive to approximations made in the determination of the longitudinal modes' wavenumbers considering the dispersive nature of waveguide modes. For the sake of comparison with the analytic model, we apply it here for an ideal beam, in order to benefit from the scaling and transparency advantages of the analytic model. Of course, final design of the TES setup should preferably be done by using the numerical simulation model [19] [18].

We first show the field waveform evolution along the undulator as predicted by the numerical computation. These were computed for the parameters of Table 1 and displayed for the case of SR in a uniform undulator (no tapering all along) in Fig. 8 and for the case of tapered undulator (TES) in Fig. 9. We sample the waveform at four positions along the undulator: $N_w$= 5, 10, 15, 20 (including the entrance and exit half period magnets). The computation is done for an ideal beam. It is correct to compare the waveform computation to the analytical curve only for the uniform section $N_w$=5 (Fig. 6), because in the absence of saturation and trapping of the bunch the analytical expression (Eq. 43) is valid there. The computed waveform of Fig. 8a compares well with the analytical curve of Fig. 6, calculated from the universal function $g_u(t'/\tau)$ (Eq.42). The maximum amplitude at the end of the first section is $E_{max} = 13MV/m$ (also as evaluated from Eq. 44). This parameter is important for evaluating the depth of the trap at the entrance to the tapered section $z_{0,2} = 0.28m$, and was used for evaluating the separatrix (Fig. 7) using (46) (53).

It is instructive to observe the scaling of the waveform evolution along the undulator in both cases of uniform undulator - SR (Fig. 8) and tapered - TES undulator (Fig. 9). The number of spikes in the wavepacket waveform grows with $N_w$ in both cases, and so is the duration of the pulse – indicating second order slippage effect (forward and backward) in the zero-slippage regime. This is consistent with the narrowing of the frequency bandwidth $\propto N_w^{-1/2}$ according to (32) (33).

In the uniform undulator case (no tapering), Fig. 8 indicates some saturation effect of the field towards the end of the undulator, mostly in the high frequency tail of the waveform. The enhanced beam energy extraction in the tapering case (Fig. 9) seems to come into expression in field enhancement of the amplitude and the high frequency tail.

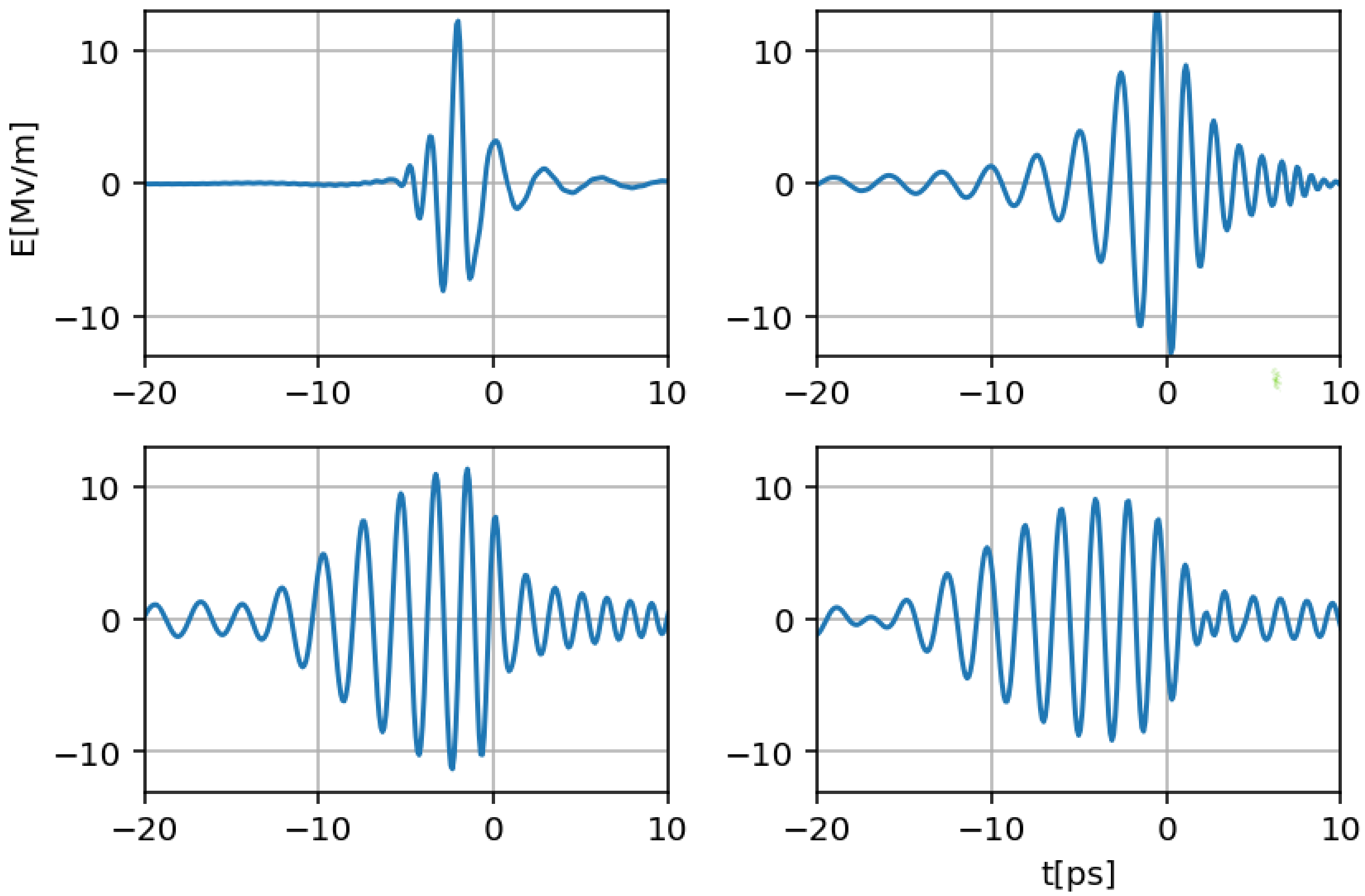

**Fig. 8** *Computed field waveforms along a uniform (untapered) undulator at positions* $N_w$*=5. 10, 15, 20 (the number of longitudinal modes for GPTFEL is 600).*

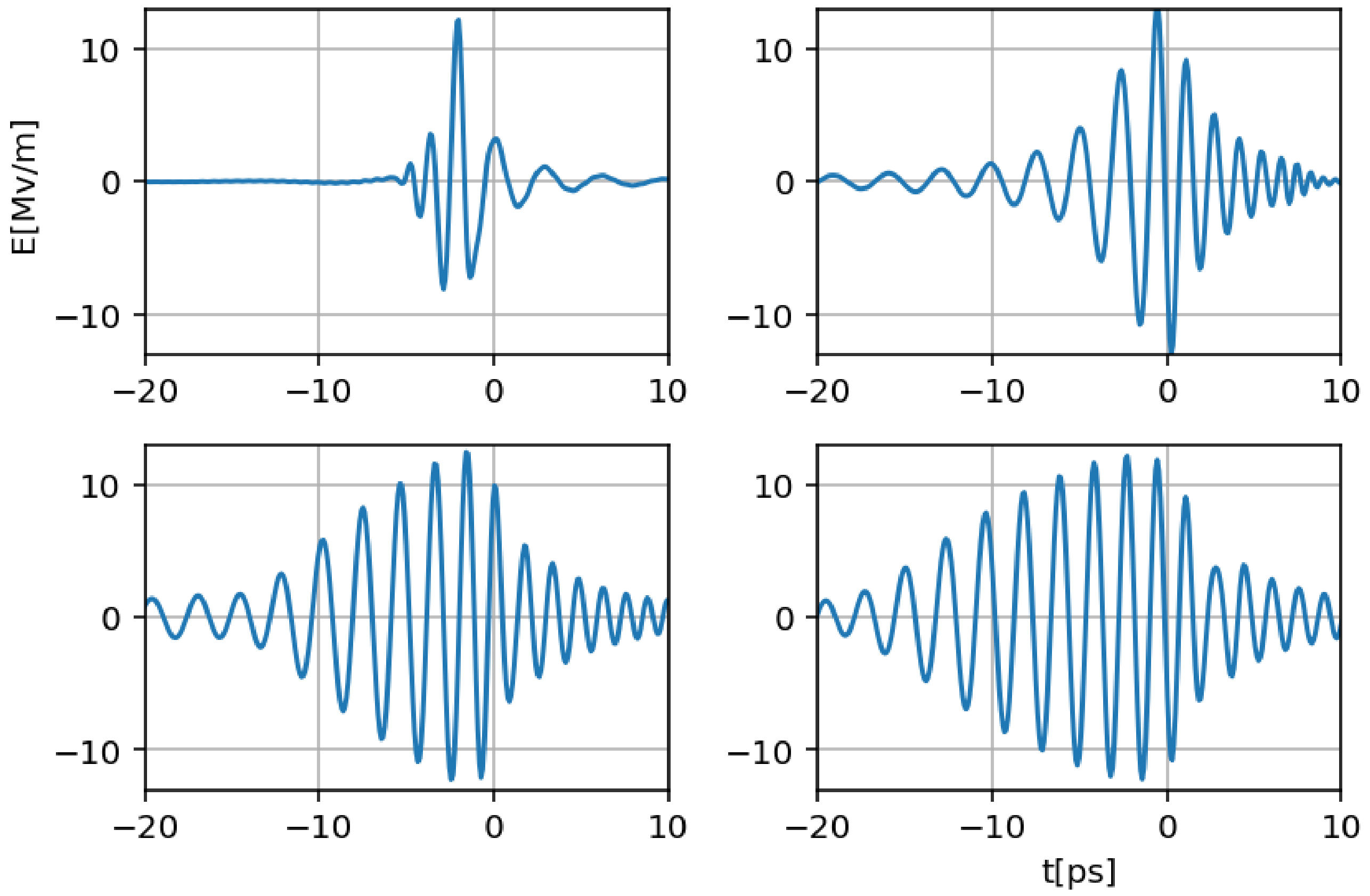

*Fig. 9: Computed field waveform along the TES (tapered)) undulator at positions* $N_w$ *=5. 10, 15, 20 (the number of longitudinal modes for GPTFEL is 600).*

The scaling of the field of the radiation wavepacket with $N_w$ is shown more transparently in Fig. 10 that shows "snapshots" of the waveforms after each 0.1 m along the entire length of the undulator. In (a) – the TES undulator (uniform section of 5 periods and tapered section of 15 periods), in (b)) - the same undulator setup but without tapering (SR emission). The field in (a) grows with $N_w$ in the uniform section up to z02=0.28m in (the same as in (b). However, in (b) the field saturates towards the end of the undulator (due to synchrotron oscillation within the trap), and in (a) the field is enhanced by approximately 25% due to the TESSA process of the tapered section.

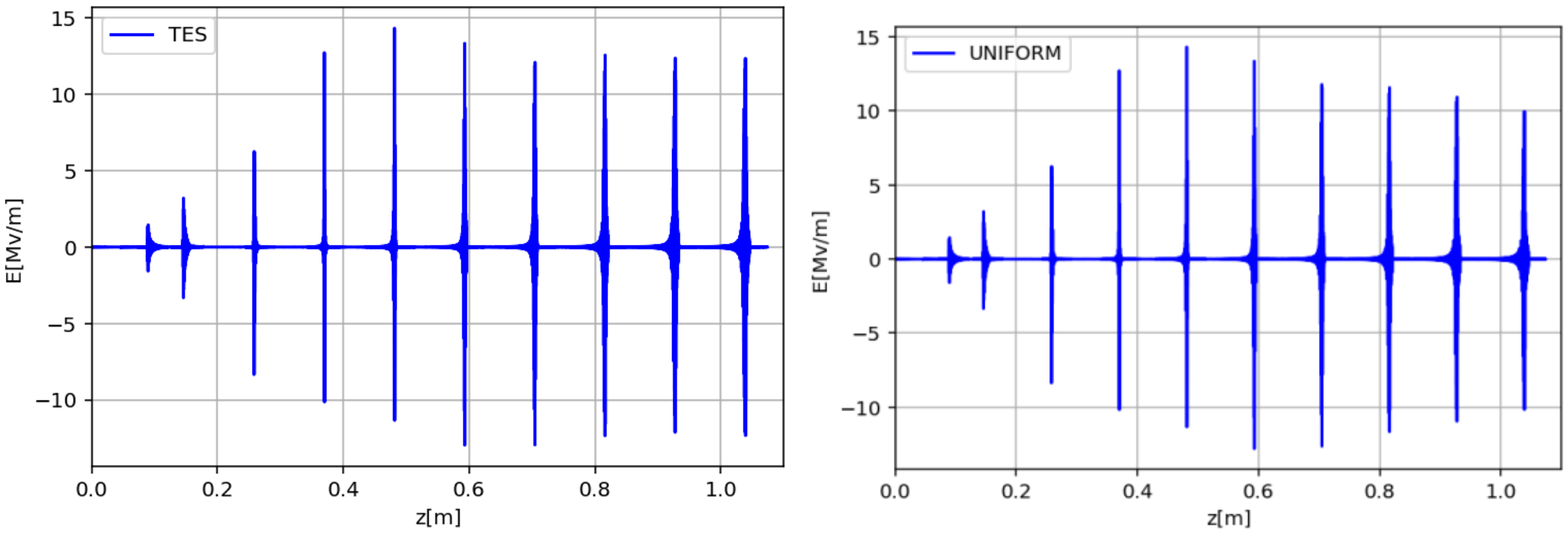

**a** **b**

Fig. 10: Snapshots of the field waveform of the radiation wavepacket along the undulator. (a) The TES (tapered) undulator. (b) A uniform undulator all along.

We now study the scaling of the wavepacket energy as function of $N_w$, computed numerically for the parameters of Table 1 for the case of TES (with tapering), and compare it to the case of SR (uniform undulator without tapering), and also compare it to the analytical estimates. This is displayed on Fig. 11 for the TES setup (uniform for $N_w = 5$ periods and tapered for the rest 15 periods) (red curve) and for an alternative undulator, uniform for the entire length (green curve). This is an instructive display, showing that without tapering the radiation energy saturates (due to synchrotron oscillation in the trap). The tapering provides enhancement of the energy by almost 100% over the saturated SR radiation in the uniform undulator owing to the TESSA process. The energy in the tapered section grows up consistently with the enhancement of the field shown in Fig. 9d.

Besides the computed curves, we show in Fig. 11 also comparison to the analytical expressions that were derived in section 3 (Eqs. 27, 28) and in section 6 (Eq. 57). The analytical expressions cannot be combined to give a good match with the numerical computations, primarily because the analytical model for the SR (section 3) is not valid in the nonlinear regime when the bunch gets trapped, and the transition from the SR process to the TESSA process does not take place at a sharp point. In Fig. 11 we use the SR expression (Eqs. 27, 28) in the uniform period section (0<z<z02=0.28m), and in the tapered section (z02<z<1.12m) we use the TESSA expression (Eq. 57) evaluate for $\psi_r = 45°$ . The analytical expression curve in the uniform section (yellow) and in the tapered section (dashed green) are slightly higher (by 20%) relative to the numerical computation curve (red).

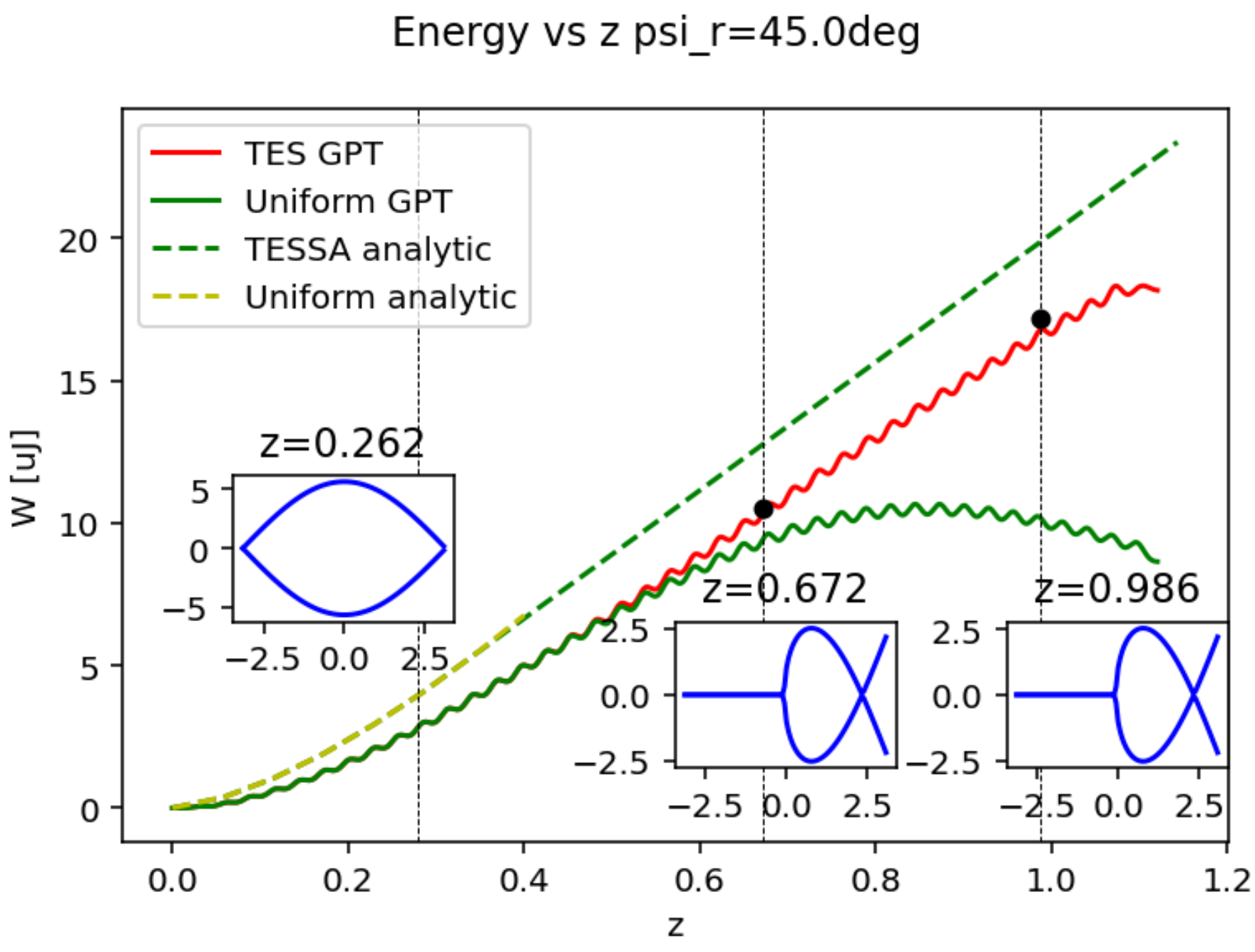

*Figure 11: The numerically computed output Energy of the radiation wavepacket in the TES (tapered undulator) configuration (red), compared to a case of an undulator uniform all along (green). Also shown is the result of the analytical model for SR (Eqs. 28, 29) in the first section (yellow). The energy growth in the second (tapered undulator) section (dashed green), is shown based on the analytical expressions Eq. 47 for* $\psi_r$ *=45°. The evolution of the phase-space separatrix is shown at various locations along the TES undulator (blue).*

## 8. Conclusions

We presented an analytical formulation for studying the radiative energy extraction and time domain field waveform evolution in TES (Tapering Enhanced Superradiance) in the zero-slippage condition of a waveguide mode. This model was used in comparison to numerical computations using FEL-GPT code.

The analytical model in the uniform undulator section of the TES setup provides estimates of the radiative energy growth and waveform development along the uniform undulator section but is valid only below saturation level (electron trapping regime). The analytical model for the tapered undulator section is based on analysis in energy-time phase-space of an electron beam bunch deeply trapped in the ponderomotive potential trap of the beam generated SR radiation. It can provide scaling laws for the energy transfer from the beam to the radiation field and useful criteria for trapping the beam, depending on the length of the undulator and the partition of the SR and TESSA sections lengths. It cannot provide expressions for the field waveform evolution.

The analytical formulation has quite good agreement with the more accurate and general numerical computation in the two parts of the TES setup. It provides transparent scaling laws that are useful for preliminary design of a TES radiation source. Evidently, the final design, including nonideal beam parameters and space-charge effect, must be done by use of the numerical computation code.

**Acknowledgement**

The author(s) declare financial support was received for the research, authorship, and/or publication of his article. We acknowledge support by the Israel Science Foundation through grant 1705/22.

**Appendix A: The spectral energy and field of Superradiant radiation**

In the case of planar undulator the transverse velocity is:

$$v_{\perp e} = \Re\{\tilde{v}_{w0} e^{ik_w z}\} = \boldsymbol{v}_{w0} cosk_W z = \frac{1}{2}\boldsymbol{v}_{w0}(e^{ik_w z} + e^{-ik_w z}) \qquad (A1)$$

Substituting in the expression for the single electron spectral work function (16):

$$\Delta\breve{W}_{qe} = -\frac{e}{v_{z0}}\int_0^{L_w} \boldsymbol{v}_{\perp\boldsymbol{e}}(z)\cdot\tilde{\boldsymbol{\epsilon}}_{\boldsymbol{q}}^{\,*}(r_{\perp e}) e^{i\left(\frac{\omega}{v_{z0}}-k_{zq}\right)} dz \qquad (A2)$$

and keeping only the resonant term, one obtains

$$\int_0^{L_w} e^{-ik_w z - ik_{zq}z + i\underline{\omega} z/v_{z0}} dz = e^{i\theta L_w}\frac{\sin\left(\frac{\theta L}{2}\right)}{\theta L/2} = L_w e^{\frac{i\theta L_w}{2}} sinc\left(\frac{\theta L_w}{2}\right)$$

$$\Delta\breve{W}_{qe} = -e\frac{\boldsymbol{v}_{\mathrm{w}0}\cdot\tilde{\boldsymbol{\epsilon}}_{\boldsymbol{q}}^{\,*}(r_{\perp e})}{2v_{z0}} L_w sinc\left(\frac{\theta L_w}{2}\right) e^{\frac{i\theta L_w}{2}} \qquad (A3)$$

where we define the detuning parameter:

$$\theta = \frac{\omega}{v_z} - k_z - k_w \qquad (A4)$$

$$v_w = \frac{c|a_w|}{\gamma} = \frac{cK}{\gamma}$$

$$\Delta\breve{W}_{qe} = -\frac{1}{2}\frac{eL_w|a_w|}{\beta_z\gamma} sinc\left(\frac{\theta}{2}L_w\right) e^{\frac{i\theta L_w}{2}} \tilde{\boldsymbol{\epsilon}}_{\boldsymbol{q}}^{\,*}(r_{\perp e}) \qquad (A7)$$

The mode amplitude is (17):

$$\check{C}_q(L_w,\omega) = \frac{1}{2P_q}\Delta\breve{W}_{qe} N M_b e^{i\omega t_0} \qquad (A8)$$

$$\check{C}_q(L_w,\omega) = -\frac{NM_b}{4P_q}\frac{eL_w|a_w|}{\beta_z\gamma}\tilde{\boldsymbol{\epsilon}}_{\boldsymbol{q}}^{\,*}(r_{\perp e}) sinc\left(\frac{\theta L_w}{2}\right) e^{\frac{i\theta L}{2}} e^{i\omega t_0} \qquad (A9)$$

We use (A9) to calculate the spectral energy. According to Parseval theorem (11):

$$\frac{dW_q}{d\omega}=\frac{1}{2\pi}P_q\left|\check{C}_q(z,\omega)\right|^2 \tag{A10}$$

Substitute (A9) in (A10):

$$\frac{dW_q}{d\omega}=\frac{N^2|M_b|^2}{2\pi P_q}\frac{1}{16}\frac{e^2L_w^2|a_w|^2}{\beta_z^2\gamma^2}|\tilde{\boldsymbol{\epsilon}}_{\boldsymbol{q}}(r_{\perp e})|^2 sinc^2\left(\frac{\theta L_w}{2}\right) \tag{A11}$$

We present the normalization power in terms of the effective area of the mode, defined by

$$\frac{1}{2Z_q}\left|\tilde{\boldsymbol{\epsilon}}_{\boldsymbol{q}}(r_{\perp e})\right|^2 A_{emq}=P_q \tag{A13}$$

Then:

$$\frac{dW_q}{dw}=\frac{e^2}{16\pi}\frac{N^2M_b^2Z_q|a_w|^2}{(\beta_z\gamma)^2}\frac{L^2}{A_{emq}}sinc^2(\frac{\theta(\omega)L_w}{2}) \tag{A13}$$

In agreement with [1]

We now go ahead to calculate the SR field. Assuming interaction with a single mode in (6):

$$\check{\mathrm{E}}(\boldsymbol{r},\omega)=\check{\mathrm{C}}_{\mathrm{q}}(z,\omega)\tilde{\boldsymbol{\epsilon}}_{\boldsymbol{q}}(r_{\perp})e^{ik_{qz}z} \tag{A14}$$

Substitute (A9):

$$\check{E}(\boldsymbol{r},\omega)=-\frac{NM_b}{4P_q}\frac{eL_w|a_w|}{\beta_z\gamma}\tilde{\boldsymbol{\epsilon}}_{\boldsymbol{q}}{}^*(r_{\perp e})\tilde{\mathcal{E}}_q(r_{\perp})e^{ik_{qz}z}sinc\left(\frac{\theta L_w}{2}\right)e^{\frac{i\theta L_w}{2}}e^{i\omega t_0} \tag{A15}$$

$$\check{E}(\boldsymbol{r},\omega)=-\frac{NM_b}{4P_q}\frac{eL_w|a_w|}{\beta_z\gamma}\left|\tilde{\boldsymbol{\epsilon}}_{\boldsymbol{q}}(r_{\perp e})\right|^2\frac{1}{\tilde{\boldsymbol{\epsilon}}_{\boldsymbol{q}}(r_{\perp e})}\tilde{\boldsymbol{\epsilon}}_q(r_{\perp})e^{ik_{qz}z}sinc\left(\frac{\theta L_w}{2}\right)e^{i\theta L_w/2}\,e^{i\omega t_0} \tag{A16}$$

In terms of effective mode area (A12):

$$\check{E}(\boldsymbol{r},\omega)=-\frac{Z_q}{2A_{em}}\frac{eL_w|a_w|}{\beta_z\gamma}\frac{NM_b}{\tilde{\boldsymbol{\epsilon}}_{\boldsymbol{q}}(r_{\perp e})}\tilde{\boldsymbol{\epsilon}}_{\boldsymbol{q}}(r_{\perp})e^{ik_{qz}z}sinc\left(\frac{\theta L_w}{2}\right)e^{i\theta L_{w}/2}\,e^{i\omega t_0} \tag{A17}$$

## Appendix B: Second order Taylor expansion of the mode dispersion curve and $\boldsymbol{\theta(\omega)}$:

We expand the mode wavenumber dispersion (3) to second order around the synchronization frequency:

$$k_z(\omega)=k_z(\omega_0)+\frac{dk_z}{d\omega}|_{\omega_0}(\omega-\omega_0)+\frac{1}{2}\frac{d^2k_z}{d\omega^2}(\omega-\omega_0)^2=k_{z0}+\frac{1}{v_g}(\omega-\omega_0)+\frac{1}{2}D(\omega-\omega_0)^2 \tag{B1}$$

where $v_g = \frac{1}{k_z'(\omega_0)}$ is the group velocity of the waveguide mode and $D = k_z''$.

Substituting (B1) into the detuning parameter $\theta(\omega) = \frac{\omega}{v_z} - k_{zq}(\omega) - k_w$ (21):

$$\theta(\omega) = \frac{\omega_0}{v_z} - k_z(\omega_0) - k_w + \left[\frac{1}{v_z} - \frac{1}{v_g}\right](\omega - \omega_0) - \frac{1}{2}D(\omega - \omega_0)^2 \qquad (B2)$$

Since at zero slippage condition (see Fig. 1) we have both phase synchronism $\theta(\omega_0) = \frac{\omega_0}{v_z} - k_z(\omega_0) - k_w = 0$ and group velocity matching $v_g = v_z$, the first and second terms vanish. We remain with

$$\theta(\omega) = -\frac{1}{2}D(\omega - \omega_0)^2 \qquad (B3)$$

In order to find the parameter D, we differentiate twice the dispersion relation (2)

$$k_z = \sqrt{(\frac{\omega}{c})^2 - (\frac{\omega_{co}}{c})^2} \qquad (B4)$$

$$D = k_z''(\omega_0) = \frac{\omega_{co}^2}{c(\omega_0^2 - \omega_{co}^2)^{\frac{3}{2}}} = \frac{1}{c^4}\frac{\omega_{co}^2}{k_{z0}^3} \qquad (B5)$$

Using the general waveguide relation: $v_{ph}v_g=c^2$, $k_z = \frac{\omega}{v_{ph}} \Rightarrow k_z = \frac{\omega v_g}{c^2} = \frac{\omega \beta_g}{c}$

Using the zero-slippage condition again ($v_g=v_z$), we arrive at:

$$D = \frac{1}{c^4}\frac{\omega_{co}^2}{k_z^3} = \frac{\omega_{co}^2}{c(\omega_0\beta_g)^3} = \frac{\omega_{co}^2}{c(\omega_0\beta_z)^3} \qquad (B6)$$

## Appendix C: Derivation of Radiated Electric field waveform for zero-slippage

To derive the radiation field in time domain for the case of zero-slippage, we use (36, 39):

$$E(\boldsymbol{r}_{\perp 0}, t) = Af_q(t) \qquad (C1)$$

$$f_q\left(t - t_0 - \frac{L_w}{v_z}\right) = \Re\left\{\frac{1}{2\pi}\int_0^\infty e^{-i\omega\left(t - t_0 - \frac{L_w}{v_z}\right)} e^{\frac{i\theta L_w}{2}} sinc\left(\frac{\theta L_w}{2}\right) e^{ik_z L_\omega} d\omega\right\} \qquad (C2)$$

and insert in it the Taylor expansion of $k_z(\omega)$ and $\theta(\omega)$ (B1, B2)

$$f_q(t) = \frac{1}{2\pi}\Re\{\int_0^\infty e^{-i\omega(t-t_0)} e^{-i\frac{D}{4}(\omega-\omega_0)^2 L_w} sinc\left(\frac{1}{4}D(\omega - \omega_0)^2 L_w\right) e^{ik_{z0}L_w + \frac{i}{vg}(\omega-\omega_0)L_w + \frac{i}{2}D(\omega-\omega_0)^2 L_w} d\omega\right\} \qquad (C4)$$

After some algebraic simplification:

$$f_q(t) = \frac{1}{2\pi}\Re[e^{ik_{z0}L_w}e^{-i\omega_0(t-t_0)}\int_0^{\infty} e^{-i(\omega-\omega_0)\left(t-t_0-\frac{L_w}{v_g}\right)} e^{+i\tau^2(\omega-\omega_0)^2} sinc(\tau^2\,(\omega-\omega_0)^2)d(\omega-\omega_0)\}(C5)$$

where we defined $\tau^2 = \frac{DL_w}{4}$.

Substituting $v_g = v_{z0}$ and using the synchronism condition $k_{z0} = \frac{\omega}{v_{z0}} - k_w$,

$$f_q(t) = \frac{1}{2\pi}\Re\{e^{-i\omega_0(t-t_0-L_w/v_{z0})}\int_0^{\infty} e^{-i(\omega-\omega_0)\left(t-t_0-\frac{L_w}{v_{z0}}\right)} e^{+i\tau^2(\omega-\omega_0)^2} sinc(\tau^2\,(\omega-\omega_0)^2)d\omega\}$$

$$f_q(t) = \frac{1}{2\pi}\Re\{e^{-i\omega_0 t'}\int_0^{\infty} e^{-i(\omega-\omega_0)t'}\, e^{+i\tau^2(\omega-\omega_0)^2} sinc(\tau^2\,(\omega-\omega_0)^2)d\omega\} \qquad (C6)$$

This represents the time dependence of the wavepacket waveform relative to the time it arrives at the end of the undulator: $t' = t - t_0 - \frac{L_w}{v_g}$ . Using $\omega' = \omega - \omega_0$ we arrive at

$$f_q(t) = \frac{1}{2\pi}\Re\left\{e^{-i\omega_0 t'}\int_{-\omega_0}^{\infty} e^{i\tau^2\omega'^2} sinc(\tau^2\,\omega'^2)d\omega']\right\} \qquad (C7)$$

With change of variables $\omega'\tau = u$,

$$f_q(t') = \frac{1}{\tau}Re\left\{e^{-i\omega_0 t'}g\left(\frac{t'}{\tau}\right)\right\} \qquad (C8)$$

where $g(\frac{t'}{\tau})$ is a dimensionless complex waveform function:

$$g\left(\frac{t'}{\tau}\right) = \frac{1}{2\pi}\int_{-\omega_0\tau}^{\infty} e^{-iu\frac{t'}{\tau}} sinc(u^2)e^{-iu^2}du \qquad (C9)$$

For narrow enough relative bandwidth of the radiation $\Delta\omega/\omega_0$, the integration can be carried out from $-\infty$ to $\infty$ and this function can be presented in terms of a universal function:

$$g_u\left(\frac{t'}{\tau}\right) = \frac{1}{2\pi}\int_{-\infty}^{\infty} e^{-iu\frac{t'}{\tau}} sinc(u^2)e^{-iu^2}du \qquad (C10)$$

Then the time dependence of the field on the beam axis upon exiting the undulator can be expressed in terms of this complex universal function as:

$$E(r_{\perp 0}, L_w, t) = Af_q(t') = \frac{A}{\tau}\left|g_u\left(\frac{t'}{\tau}\right)\right|\cos\left(\omega_0 t' - \phi_g\left(\frac{t'}{\tau}\right)\right) \qquad (C11)$$

where $|g_u(t'/\tau)|$ and $\phi_g(t'/\tau)$ are the amplitude and phase of $g_u(t'/\tau)$ respectively. These are displayed in Fig. 5a and b respectively. In Fig. 5 we also overlay the curve of equation (41), calculated for $\omega_0\tau = 2.36$, corresponding to the case of $N_w = 5$. It shows that the universal complex function (42) is a good approximation even in this wide bandwidth case (Fig. 3a).

The universal function $g_u(x) = g_u(t'/\tau)$ can be presented in an analytical form:

$$g_u(x) = \left[(1-i)e^{\frac{ix^2}{8}}\sqrt{\pi} + \frac{1}{2}\pi i\left(|x| - (1-i)x\left[C\left(\frac{x}{2\sqrt{\pi}}\right) + iS\left(\frac{x}{2\sqrt{\pi}}\right)\right]\right)\right] \qquad (C12)$$

C, S being the Fresnel Integrals.

$S\left(\frac{t}{2\sqrt{\pi}}\right)$